\documentclass[twocolumn,amsmath,showpacs]{revtex4}
\usepackage{graphicx}

\begin{document}

\title{Synchronous Optical Pumping of Quantum Revival Beats for Atomic Magnetometery}
\author{S. J. Seltzer}
\author{P. J. Meares}
\author{M. V. Romalis}

\begin{abstract}
We observe quantum beats with periodic revivals due to non-linear
spacing of Zeeman levels in the ground state of potassium atoms and
demonstrate their synchronous optical pumping by double modulation
of the pumping light at the Larmor frequency and the revival
frequency. We show that synchronous pumping increases the degree of
spin polarization by a factor of 4. As a practical example, we
explore the application of this double-modulation technique to
atomic magnetometers operating in the geomagnetic field range and
find that it can increase the sensitivity and reduce magnetic field
orientation-dependent measurement errors endemic to alkali-metal
magnetometers.
\end{abstract}

\affiliation{Department of Physics, Princeton University, Princeton,
New Jersey 08544}


\maketitle

Periodic revival of quantum beats is a general phenomenon that
occurs in multi-level systems with non-linear energy level spacing
\cite{Leichtle,Robinett} and has been observed in diverse systems,
from one-atom masers \cite{Walther} and Rydberg states in atoms
\cite{Stroud} to molecular vibrational \cite{Bowman} and rotational
\cite{Rosca} states. Recently molecular rotational revivals
attracted significant attention because of the possibility of using
them for alignment of molecules with ultrashort laser pulses
\cite{Seideman,Stapelfeldt,Bisgaard,Daems}. It has been proposed
that periodic laser pulses could be used to increase the degree of
molecular alignment and maintain it indefinitely
\cite{Leibscher,Leibscher2,Ortigoso,Ortigoso2}. An increase of
molecular alignment using two pulses has been demonstrated in
\cite{Stapelfeldt1}.

Here we show experimentally that appropriately synchronized train of
laser pulses can increase in the degree of spin orientation and
maintain it indefinitely by synchronously pumping quantum revivals
in the ground-state Zeeman levels of an alkali-metal atom. Periodic
revivals of quantum beats in Zeeman levels of alkali-metal atoms
occur naturally due to non-linear Breit-Rabi mixing
\cite{Breit-Rabi} and have been modeled in \cite{AlexBeats}. We
observe quantum revivals in K atoms and demonstrate their
synchronous pumping by double modulation of the pump laser at both
the Larmor and the revival frequencies. We find that it creates a
coherently oscillating superposition state with a spin polarization
a factor of 4 higher than can be obtained without double modulation
for the same average laser power. We also model the effects in other
alkali atoms and find that the amount of polarization enhancement
increases for systems with a larger number of quantum states.

Because the revivals occur in geomagnetic field range of about
0.5~G, our experiments are also directly applicable to
optically-pumped alkali-metal magnetometers,  which are used in many
applications, from archaeology \cite{David} and mineral exploration
\cite{Nabighian} to searches for a CP-violating electric dipole
moment \cite{Groeger}. As was recently discussed in
\cite{Kitching2,Acosta}, Breit-Rabi mixing of Zeeman levels
decreases magnetometer sensitivity by splitting the Zeeman resonance
into many separate lines. We show that synchronous pumping of
quantum revivals largely recovers the signal loss. In addition, we
demonstrate that it generates a symmetric resonance lineshape,
reducing a systematic effect endemic to most alkali-metal
magnetometers, known as a heading error, that causes changes in the
measured absolute value of the magnetic field depending on the
orientation of the magnetometer with respect to the magnetic field
vector \cite{AlexHead}. We expect that synchronous pumping of
revivals can be also used to improve the precision and accuracy of
spectroscopic measurements on other multi-level systems.

For alkali-metal atoms with nuclear spin $I$, the energy levels of
the two hyperfine states with $F=I\pm 1/2$ in a magnetic field $B$
are given by well-known Breit-Rabi equation \cite{Breit-Rabi}.
Following the general description of quantum revivals in multi-level
systems \cite{Robinett}, the energy levels can be expanded in powers
of $m_{F}$ keeping only the leading $B$ dependence in each term,
\begin{eqnarray}
E(F,m_{F}) &=&-\frac{h\nu _{\mathrm{hf}}}{2(2I+1)}+\left( -g_{I}\mu
_{N}\pm
\mu _{\rm eff}\right) Bm_{F}  \nonumber \\
&&\mp \frac{\mu _{\rm eff}^{2}B^{2}m_{F}^{2}}{h\nu
_{\mathrm{hf}}}\pm 2\frac{\mu _{\rm eff}^{3}B^{3}m_{F}^{3}}{(h\nu
_{\mathrm{hf}})^{2}},
\end{eqnarray}
where $\mu _{\rm eff}=(g_{s}\mu _{B}+g_{I}\mu _{N})/(2I+1)$, $h\nu
_{\mathrm{hf}}$ is the hyperfine splitting of the ground state,
$g_{I}=\mu_{I} /(\mu _{N}I)$ is the nuclear g-factor, $\mu_{I}$ is
the nuclear magnetic moment, $g_{s}\simeq 2$ is the electronic
g-factor, $\mu _{N}$ is the nuclear magneton, and $\mu _{B}$ is the
Bohr magneton. The Larmor frequency corresponds to classical spin
precession $\nu _{L}=\left(
-g_{I}\mu _{N}\pm \mu _{\rm eff}\right) B/h$, the revival frequency is given by $%
\nu _{\mathrm{rev}}=\mu _{\rm eff}^{2}B^{2}/h^{2}\nu _{\mathrm{hf}}$
and the super-revival frequency is $\nu _{\mathrm{suprev}}=2\mu
_{\rm eff}^{3}B^{3}/h^{3}\nu _{\mathrm{hf}}^{2}$. For $^{39}$K atoms
$(I=3/2, g_{I}=0.26, \nu _{\mathrm{hf}}=461.7$ MHz)
in a magnetic field $B=0.5$ G we have $\nu _{L}=350$ kHz, $\nu _{%
\mathrm{rev}}=265$ Hz and $\nu _{\mathrm{suprev}}=0.4$ Hz.


The experiments are performed in a 5 cm diameter evacuated spherical
Pyrex cell with a 1 mm diameter, 5 cm long stem containing K metal
in natural abundance (93$\%\ ^{39} $K and 7$\%\ ^{41}$K). The cell
is heated to 60\ensuremath{^{\circ}}C by hot air in a double-walled
glass oven. The cell walls are coated with octadecyltrichlorosilane
(OTS) to reduce surface spin relaxation of K atoms. OTS molecules
(CH$_{3}$-(CH$_{2}$)$_{17}$-Si-Cl$_{3}$) attach to glass through a
silanization reaction and ideally expose alkali atoms only to a
surface of long hydrocarbon chains, known to reduce wall relaxation
due to their low polarizability \cite{Bouchiat}. We have used the
coating procedure described in \cite{Rosen} and obtained K
longitudinal spin relaxation time T$_{1}$ up to 145 msec,
corresponding to 2100 collisions with the walls before
depolarization, although the relaxation times were not entirely
consistent and other cells coated in the same batch had T$_{1}$ of
1.6 ms, 20 ms, and 61 ms. The longitudinal spin relaxation time did
not degrade and actually improved in some cells when they were
heated to 120\ensuremath{^{\circ}}C for a period of weeks. In
magnetic resonance measurements we found that $T_{2}$ was limited by
relaxation due to magnetic field gradients to about 30 msec.

\begin{figure}[tbp]
\centering
\includegraphics[width=0.9\columnwidth]{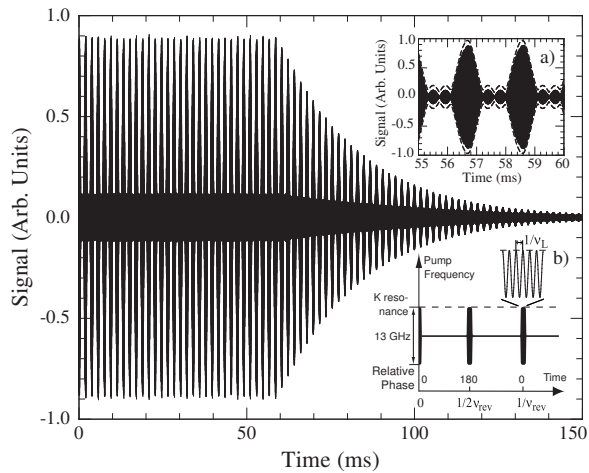}
\caption{Measurement of $\langle S_x \rangle$ showing multiple
quantum revivals. Synchronous  pumping of quantum revivals is
started 150 msec before $t=0$ and stopped at $t=60$ msec, allowing
free decay of spin coherence.  Inset a) shows a comparison of the
revival envelope with a density matrix simulation (dashed lines),
while inset b) shows double-modulation of the pump laser frequency
so the optical pumping only occurs during instances of maximum
expectation value of $\langle S_z \rangle$.}\label{fig_time}
\end{figure}

In most measurements we use the Bell-Bloom technique \cite{Bell2} of
spin resonance excitation using a pump laser modulated at the Larmor
frequency. The cell is placed in three orthogonal Helmholtz coils
which are used to control the magnitude and direction of the
magnetic field, as well as five gradient coils used to cancel
ambient field gradients. The field is nominally set along the
$\hat{y}$ direction, and magnetic field noise along this direction
is actively canceled using feedback from a fluxgate sensor
(Bartington Instruments Mag-03MC) located next to the oven.  The
atoms are pumped by circularly-polarized light from a DFB diode
laser propagating along the $\hat{z}$ direction. The laser is
detuned by approximately 6.3 GHz off the potassium D1 resonance at
770.1 nm, much larger than the measured 0.9 GHz Doppler-broadened
halfwidth of the optical line. The laser current is sinusoidally
modulated at the Larmor frequency so that the laser frequency
reaches the resonance only for a short fraction of each cycle.  For
synchronous pumping of quantum revivals a secondary modulation is
applied at twice the revival frequency, turning the Larmor frequency
current modulation on and off with a duty cycle in the range of
1-10\%, as shown in Fig.~\ref{fig_time}b. In addition, the relative
phase of the Larmor modulation is changed by
180\ensuremath{^{\circ}}\ for odd modulation pulses. In frequency
space this creates sidebands of the optical pumping rate that are
located at $\nu _{L}\pm \nu_{\rm rev}(2n+1),$\thinspace $n=0,1,2...$
When the high frequency modulation is tuned exactly to $\nu _{L}$
for $F=2$, the sidebands of the pumping rate modulation
simultaneously excite all $F=2$ Zeeman transitions. Simultaneous
excitation of two Zeeman coherences in an atomic magnetometers has
been previously demonstrated in \cite{Balabas}. The pump beam power
is always adjusted to broaden magnetic resonance width by a factor
of 2, which gives the same average pumping rate that approximately
maximizes the signal-to-noise ratio in atomic magnetometers.

\begin{figure}[b]
\centering
\includegraphics[width=0.8\columnwidth]{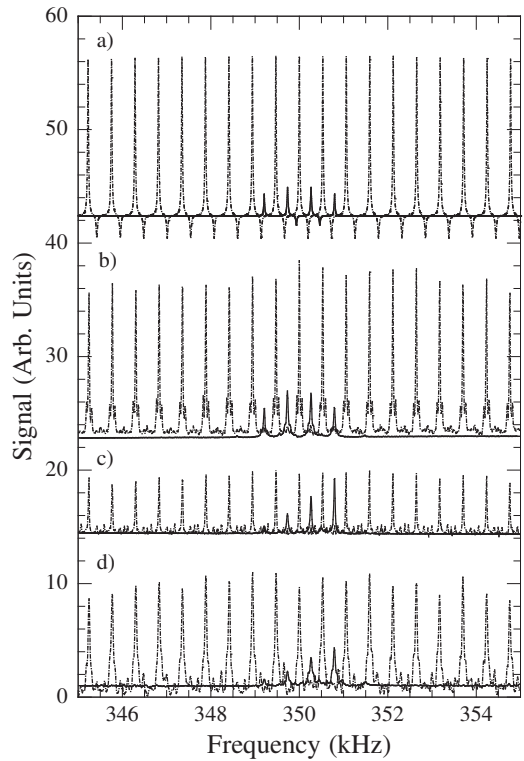}
\caption{Magnetic resonance spectra in a 0.5 G field with (dashed
lines) and without (solid lines) secondary modulation at twice the
revival frequency. Numerical simulation (panel a) and experimental
measurements with magnetic field in $\hat{y}$ direction (panel b)
and tilted by 60\ensuremath{^{\circ}}\ into the $\hat{z}$ direction
(panel c) with optical excitation of spin precession. Panel
d:``$M_x$" magnetometer signal with rf excitation and magnetic field
in $\hat{z}$ direction. The signals are offset in the vertical
direction by an arbitrary amount. Small random variation in the
height of the peaks with secondary modulation and small sidebands
are due to 60 Hz noise.} \label{fig_spectra}
\end{figure}

The $\langle S_x \rangle$ component of spin polarization  is probed
using optical rotation of a DFB laser beam propagating along the
$\hat{x}$ direction and detuned about 1 GHz away from the D1
resonance. Synchronous optical pumping of quantum revivals is shown
in Fig.~\ref{fig_time} with a 10\% duty cycle for secondary
modulation. Beating between different Zeeman coherences causes the
polarization envelope to reappear with a period $2/\nu _{rev}$. As
shown in Fig.~\ref{fig_time}a), the  shape of the revival envelope
is in agreement with a density matrix simulation. The revivals
persist indefinitely, until at $t=60$ msec the pumping light is
turned off, showing free spin precession with many quantum revivals
with amplitude decaying due to spin relaxation.

To obtain a magnetic resonance signal, the amplitude of optical
rotation at the Larmor frequency, proportional to $\langle
S_x\rangle$, is recorded by a lock-in amplifier and sampled by a
sample-and-hold circuit triggered at times of maximum revival. For
optimal detection of spin precession the probe laser can be turned
on only during revivals with a duty factor $d$ and a power increased
by a factor of $1/d$ relative to optimal continuous measurement,
giving the same shot noise sensitivity.

Figure \ref{fig_spectra} shows the magnetic resonance spectrum for
$B\simeq $ 0.5 G as a function of modulation frequency of the pump
laser current both with (dashed lines) and without (solid lines)
secondary modulation at 530 Hz with a 1\% duty cycle. Without
secondary modulation the resonance spectrum consists of 4 lines
corresponding to individual $F=2, m_{F}\rightarrow m_{F}+1$ Zeeman
transitions. With secondary modulation there is a central resonance
at $\nu _{L}$, strictly proportional to the magnetic field, and a
number of sidebands spaced by $2\nu _{\rm rev}$.  The density matrix
simulation shown in Fig.~\ref{fig_spectra}a predicts that the
maximum spin polarization is increased by a factor of 6 relative to
the case without secondary modulation due to constructive
interference between all Zeeman coherences; the experimental data
plotted in Fig.~\ref{fig_spectra}b show an increase of the spin
polarization by a factor of 3.9. The linewidth of the resonances is
kept the same so higher spin polarization directly translates into
higher magnetometer sensitivity.

The heading errors in atomic magnetometers can be simply understood
as resulting from overlap of the multiple Zeeman resonances
\cite{AlexHead}. For well-resolved resonances, as in
Fig.~\ref{fig_spectra}, the magnetometer is locked to the strongest
resonance and the heading error is due to tails of other resonances
that change in size depending on the orientation of the pump laser
relative to the magnetic field because of changes in the
longitudinal spin polarization. The frequency shift of a particular
resonance with frequency $\nu _{0}$ and amplitude $A_{0}$ due to the
tails of other Lorentizan resonances can be estimated as
\begin{equation}
\Delta \nu =\frac{\Gamma ^{2}}{A_{0}}\sum_{i}\frac{A_{i}}{\nu
_{i}-\nu _{0}}, \label{eq_shift}
\end{equation}
where $\Gamma $ is half-width at half-maximum, assumed to be the
same for all resonances, and $\nu _{i}$ and $A_{i}$ are the
frequencies and relative heights of the other resonance lines.
Magnetometer signals recorded with the magnetic field tilted by 60%
\ensuremath{^{\circ}}\ from the $y$ axis in the direction of the
pump laser are shown in Fig.~\ref{fig_spectra}c). The relative
strengths of Zeeman resonances without secondary modulation change
significantly depending on field orientation, leading to a heading
error given by Eq.~(\ref{eq_shift}). In contrast, the resonance
lines obtained with secondary modulation are symmetrical and do not
systematically change relative size with tilting of the field, so
the tails of various peaks cancel and do not lead to a significant
heading error.
\begin{figure}[b]
\centering
\includegraphics[width=0.8\columnwidth]{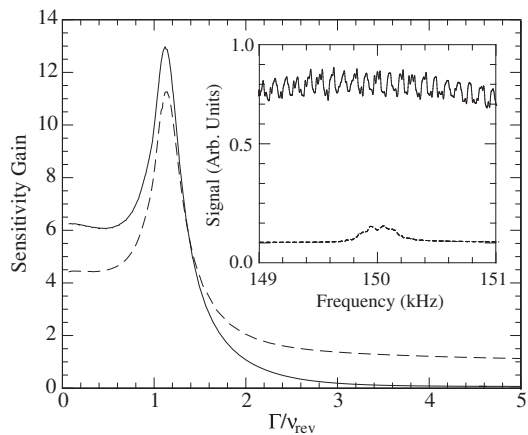}
\caption{Calculated gain in magnetometer sensitivity due to higher
spin polarization for secondary modulation with 1\% (solid line) and
30\% (dashed line) duty cycle relative to the case with no secondary
modulation. The average optical pumping rate is set equal to the
relaxation rate.  The inset shows experimental magnetic resonance
spectrum in a 0.21 G field with increased magnetic field gradients.
Secondary modulation (solid line) shows individual resonances which
are poorly resolved without it (dashed line).} \label{fig_Sens}
\end{figure}

We also point out the general applicability of the double-modulation
technique by demonstrating that it works for a common
``$M_{x}$"-type magnetometer \cite{AlexHead} using an rf field to
excite spin precession. For this data the magnetic field is directed
along $\hat{z}$, parallel to the pump laser, which is not modulated
and remains tuned to the D1 resonance. Spin precession is excited by
a transverse rf field at the Larmor frequency which is similarly
modulated at twice the revival frequency with 1\% duty cycle and 180%
\ensuremath{^{\circ}} phase reversal for odd pulses. The solid line
in Fig.~\ref{fig_spectra}d) shows continuous rf excitation while the
dashed line shows  rf excitation with secondary modulation. The
amplitude of the rf field is adjusted to generate the same level of
rf broadening, ensuring equal widths of the resonances. As with
optical excitation, we observe an increase in sensitivity by
approximately a factor of 3 and a symmetrical set of sidebands that
does not cause a significant heading error.

\begin{figure}[tbp]
\centering
\includegraphics[width=0.9\columnwidth]{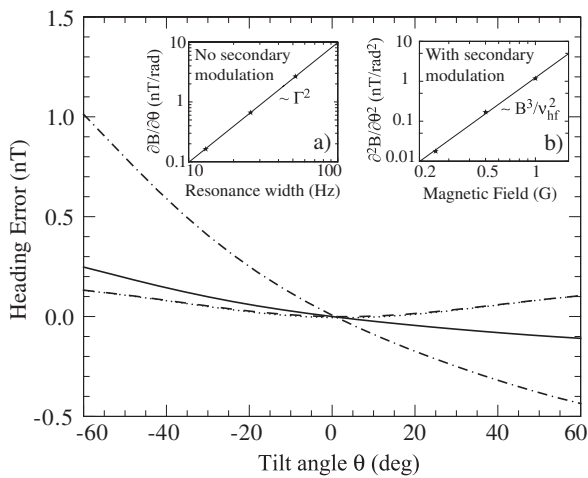}
\caption{The heading error determined from numerical simulation as a
function of the magnetic field tilt angle $\theta$ away from the $y$
axis into the direction of the pump beam. With no secondary
modulation the heading error depends on the resonance linewidth
(solid line - $\Gamma = 13$ Hz, dash-dot line - $\Gamma = 26$ Hz),
while with secondary modulation the heading error is approximately
quadratic in $\theta$ and is independent of the resonance linewdith
(dashed line - $\Gamma = 13$ Hz, dotted line - $\Gamma = 26$ Hz).
Inset a) shows that without secondary modulation the slope of the
heading error near $\theta=0$ scales as $\Gamma^2$. Inset b) shows
that with secondary modulation the curvature of the heading error
near $\theta=0$ scales as $B^3$, indicating that it is due to
third-order splitting of the energy levels. } \label{fig_Errorsig}
\end{figure}

We use a numerical simulation for a general analysis of polarization
gains and heading errors. We calculate the time evolution of the
alkali-metal ground state density matrix in the presence of
doubly-modulated pumping and simulate lock-in detection, generating
resonance spectra (see Fig.~\ref{fig_spectra}a) that are similar to
experimental data. The magnetometer sensitivity and the resonance
frequency are determined, respectively, from the slope and the
zero-crossing of the dispersion curves. Fig.~\ref{fig_Sens} shows
the gain in sensitivity due to higher spin polarization for K  as a
function of resonance linewidth. We also investigated the case of Cs
with I=7/2 and found that the increase in maximum spin polarization
and the gain in sensitivity is about a factor of 2 higher than in K
due to coherent interference of a larger number of levels. Secondary
modulation also helps to resolve individual resonances when $\Gamma
\sim \nu_{\rm rev}$, as demonstrated experimentally in the inset of
Fig.~\ref{fig_Sens}, leading to additional gain in sensitivity over
the case of continuous pumping in this regime. The heading error,
which is usually on the order of $\nu_{\rm rev}$ when the individual
resonances are poorly resolved, can also be significantly reduced by
locking  to one of the resolved resonances, which remain symmetrical
when the magnetic field is tilted.

In Fig.~\ref{fig_Errorsig} we show the  heading error as a function
of magnetic field tilt angle $\theta $ into the direction of the
pump beam. Without secondary modulation the heading error is
approximately linear near $\theta =0$ and scales with the square of
the resonance linewidth $\Gamma $, as predicted by Eq.
(\ref{eq_shift}). In contrast, the heading error with secondary
modulation is quadratic near $\theta =0$ and does not depend on the
resonance linewidth, making the magnetometer less susceptible to
magnetic gradient broadening; it is instead due to cubic energy
level splitting and scales as $B^{3}/\nu _{\mathrm{hf}}^{2}$. We
expect the heading error with secondary modulation to be much
smaller for Rb or Cs with larger $\nu_{\mathrm{hf}}$. For
comparison, most existing atomic magnetometers have heading errors
on the order of 10 nT and can achieve 0.1 nT level only with very
narrow resonance lines \cite{AlexHead}.


In conclusion, we demonstrated synchronous pumping of quantum
revivals and shown that it can be used to increase the degree of
atomic orientation.  We also find that the resulting resonance
lineshape is symmetric, which is important for reducing systematic
errors in frequency measurements on multi-level systems. Our results
are directly applicable to alkali-metal magnetometers operating in
geomagnetic field range and can be used to increase their
sensitivity and reduce heading errors.  This work was funded by an
Office of Naval Research MURI grant and the NSF.


\begin{thebibliography}{99}

\bibitem{Leichtle}  C. Leichtle, I.S. Averbukh, W.P. Schleich, Phys. Rev.
Lett. \textbf{77} 3999 (1996).

\bibitem{Robinett}  R.W. Robinett, Phys. Rep. \textbf{392}, 1 (2004).

\bibitem{Walther}  G. Rempe, H. Walther, and N. Klein, Phys. Rev. Lett. \textbf{58},
353 (1987).

\bibitem{Stroud}  J.A. Yeazell, M. Mallalieu, and C. R. Stroud, Jr., Phys.
Rev. Lett. \textbf{64}, 2007 (1990).


\bibitem{Bowman}  R.M. Bowman, M. Dantus, and A.H. Zewail, Chem. Phys. Lett.
\textbf{161}, 297 (1989).

\bibitem{Rosca} F. Rosca-Pruna and M. J. J. Vrakking, Phys. Rev. Lett. {\bf 87}, 153902
(2001).



\bibitem{Seideman} T. Seideman, Phys. Rev. Lett. {\bf 83}, 4971
(1999).

\bibitem{Stapelfeldt} H. Stapelfeldt and T. Seideman, Rev. Mod.
Phys.  {\bf 75}, 543 (2003).

\bibitem{Bisgaard} C.Z. Bisgaard {\it et. al}, Phys. Rev. Lett. {\bf
92}, 173004 (2004).

\bibitem{Daems} D. Daems {\it et. al}, Phys. Rev. Lett. {\bf
95}, 063005 (2005).

\bibitem{Leibscher} M. Leibscher, I. Sh. Averbukh, and H. Rabitz,
Phys. Rev. Lett. {\bf 90}, 213001 (2003).

\bibitem{Leibscher2} M. Leibscher, I. Sh. Averbukh, and H. Rabitz, Phys. Rev. A {\bf 69}, 013402
(2004).

\bibitem{Ortigoso} J. Ortigoso, Phys. Rev. Lett. {\bf 93}, 073001,
(2004).

\bibitem{Ortigoso2} J. Ortigoso and J. Santos, Phys. Rev. A {\bf 72}, 053401
(2005).


\bibitem{Stapelfeldt1} C.Z. Bisgaard, M.D. Poulsen, E. Péronne, S. S.
Viftrup, and H. Stapelfeldt, Phys. Rev. Lett. {\bf 92} , 173004
(2004).

\bibitem{Breit-Rabi}  G. Breit and I.I. Rabi, Phys. Rev. \textbf{38} 2082
(1931).

\bibitem{AlexBeats}  E.B. Alexandrov \textit{et al.}, J. Opt. Soc. Am. B \textbf{22}, 7 (2005).



\bibitem{David} A. David \textit{et al.}, Antiquity \textbf{78}, 341 (2004).

\bibitem{Nabighian} M.N. Nabighian \textit{et al.}, Geophys. \textbf{70} 33
(2005).

\bibitem{Groeger} S. Groeger, A. S. Pazgalev, and A. Weis, Appl. Phys.
B. \textbf{80}, 645 (2005).


\bibitem{Kitching2}  P.D.D. Schwindt, L. Hollberg and J. Kitching, Rev. Sci.
Inst. \textbf{76} 126103 (2005).

\bibitem{Acosta}  V. Acosta \textit{et al.}, Phys. Rev. A {\bf 73}, 053404 (2006)

\bibitem{AlexHead}  E.B. Alexandrov, Phys. Scripta T {\bf 105}, 27
(2003).



\bibitem{Bouchiat} M. A. Bouchiat and J. Brossel, Phys. Rev. {\bf 147}, 41
(1966).


\bibitem{Rosen}  M.S. Rosen \textit{et al.}, Rev. Sci. Instrum. \textbf{70}, 1546 (1999).


\bibitem{Bell2}  W.E. Bell and A.L. Bloom, Phys. Rev. Lett. \textbf{6}, 280
(1961).

\bibitem{Balabas} M.V. Balabas, V.A. Bonch-Bruevich, and S.V.
Provotorov, Sov. Tech. Phys.  Lett. {\bf 15}, 287 (1989).


\end{thebibliography}
\end{document}